\documentclass{aa} % 
\usepackage{graphicx,color}
\usepackage{txfonts}
\begin{document}

\title{Wind suppression by X-rays in Cygnus X-3}
  \author{O. Vilhu
         \inst{1}
          \and
          T.R. Kallman
     \inst{2}
                \and
     K.I.I. Koljonen
       \inst{3,4}
         \and
       D.C. Hannikainen
  \inst{5}
            }
 \offprints{O. Vilhu}
  \institute{Dept of Physics, P.O. Box 84, FI-00014  University of Helsinki, Finland\\
              \email{osmi.vilhu@gmail.com}
 \and
 Lab. for High-Energy Astrophysics, NASA/GSFC, Greenbelt, Maryland, USA\\       \email{timothy.r.kallman@nasa.gov}
\and
           Finnish Centre for Astronomy with ESO (FINCA), University of Turku, Väisäläntie 20, 21500 Piikkiö, Finland\\
                \email{karri.koljonen@utu.fi}
\and
          Aalto University Metsähovi Radio Observatory, PO Box 13000, Fi-00076 Aalto, Finland\\
          \email{karri.koljonen@utu.fi}
  \and
           Sky \& Telescope, One Alewife Center, Cambridge,  MA 02140, USA\\
                   \email{diana@skyandtelescope.org}
     }

   \date{Received 22.2.2021; accepted 25.3.2021  }

 %-------------------------------------------------------------------------------------------
  \abstract
{The radiatively driven wind of the primary star in wind-fed X-ray binaries can be suppressed by the X-ray irradiation of the compact secondary star. This causes feedback between the wind and the X-ray luminosity of the compact star.}
{   We aim to estimate how the wind velocity on the face-on side of the donor star depends on the spectral state of the high-mass X-ray binary Cygnus X-3.  }
{    We modeled the supersonic part of the wind by computing the line force (force multiplier) with the Castor, Abbott \& Klein formalism and XSTAR physics and by solving the mass conservation and momentum balance equations. We computed the line force locally in the wind considering the radiation fields from both the donor and the compact star in each spectral state. We solved the wind equations at different orbital angles from the line joining the stars and took the effect of wind clumping into account. Wind-induced accretion luminosities were estimated using the Bondi-Hoyle-Lyttleton formalism and computed wind velocities at the compact star. We compared them to those obtained from observations.
  }
{     We found that the ionization potentials of the ions contributing the most to the line force fall in the extreme-UV region (100-230 Å). If the flux in this region is high, the line force is weak, and consequently, the wind velocity is low. We found a correlation between the luminosities estimated from the observations for each spectral state of Cyg X-3 and the computed accretion luminosities assuming moderate wind clumping and a low mass of the  compact star. For high wind clumping, this correlation disappears. We compared the XSTAR method used here with the comoving frame method and found that they agree reasonably well with each other.
  }
  %  conclusions heading (optional), leave it empty if necessary 
{ We show that soft X-rays in the extreme-UV region from the compact star penetrate the wind from the donor star and diminish the line force and consequently the wind velocity on the face-on side. This increases the computed accretion luminosities qualitatively in a similar manner as observed in the spectral evolution of Cyg X-3 for a moderate clumping volume filling factor and a compact star mass of a few (2 - 3) solar masses.}   


   \maketitle
%
%_______________________________________________________________

\section{Introduction}

Cygnus X-3 (4U 2030+40) was one of the first X-ray binaries  discovered (Giacconi et al. 1967) and is also one of the brightest. It is listed in the Catalogue of High-Mass X-ray Binaries in the Galaxy (Liu et al. 2006), where most entries consist of an OB star whose wind feeds a companion neutron star or a black hole, releasing X-rays in the process.  The donor star of Cyg X-3 (optical counterpart V1521 Cyg) is a massive Wolf-Rayet (WR) star of either of type WN 5-7 (van Keerkwijk et al. 1992 and 1996) or a weak-lined WN 4-6 (Koljonen \& Maccarone 2017). The relatively small size of this helium star allows for a tight orbit with a period of 4.8 hours that is more typical of low-mass X-ray binaries. In this paper we study the effect of soft X-rays (EUV radiation) on various wind models of the WR star and estimate the corresponding accretion rates and luminosities during different spectral states of Cyg X-3.

Sander et al. (2017) presented three different approaches for handling the radiative transfer of line-driven winds of hot stars. The first is an analytical description based on the concept of the CAK theory (Castor, Abbott \& Klein 1975). This method has also opened up the whole area of time-dependent and multidimensional calculations (see, e.g., Sundqvist et al. 2010). The CAK theory assumes that line broadening is dominated by the bulk motion of the wind. In a rapidly expanding wind, line saturation is avoided, which leads to a strong line force. The second approach is to calculate the radiative force with the help of Monte Carlo (MC) methods. Computationally, the MC method is much more time-consuming than the CAK method. The third approach is to calculate the radiative transfer in the comoving frame (CMF; see Krticka \& Kubat 2017; Krticka et al. 2018; Sander at al. 2018). This method has also been applied to WR stars (Sander et al. 2017; Hamann et al. 2006; Hainich et al. 2014).

X-rays (mainly soft X-rays in the extreme-UV, EUV, region) from the compact star affect the wind through photoionization. An increase in highly ionized ions may decrease the line-driven mechanism and reduce the wind velocity. As noted by Karino (2014), pioneering work on these ionized wind dynamics was done by Stevens and Kallman (1990), who used the XSTAR code (Kallman 2018) to approximate the wind with a photoionized plasma; the line force was computed with the CAK method.
In the present paper, we revisit and further develop this method and apply it to Cyg X-3. We adopt the first approach (CAK) classified by Sander et al. (2018).  The results we present agree with those of Krticka et al. (2018) for HMXRBs: X-rays reduce the wind velocity, and wind clumping weakens the effect of X-rays. Szostek and Zdziarski (2008) found that the wind of Cyg X-3 has to be clumpy in order to explain the observed X-ray  absorption. We therefore include clumping with different filling factors here. 

Our goal is to study the penetration of X-rays from the compact star of Cyg X-3 of the wind, and to determine how it affects the wind properties, in particular, the wind velocity between the two stars. Based on this, we estimate the accretion luminosities of the five main spectral states of Cyg X-3 as classified by Hjalmarsdotter et al. (2009). MacGregor and Vitello (1982) applied a somewhat similar approach to X-ray binary systems with wind accretion and found that X-ray irradiation diminishes the wind velocity. A relevant recent paper (and references therein) is on Cyg X-1 by Meyer-Hofmeister et al. (2020). They found that the outflow from the OB-star facing the X-ray source is noticeably slower than from the opposite hemisphere in the X-ray shadow. Furthermore, they give evidence for two stable states: the soft high-luminosity state, and the hard low-luminosity state.

\section{Cyg X-3 system parameters }
Using Chandra High Energy Transmission Grating data (HETG, ObsID 7268 and 6601), Vilhu et al. (2009) derived radial velocity curves of the FeXXVI emission line at 1.78 Å (H-like), which is thought to originate in the vicinity of the compact star. These data comprise ten phase bins and give a radial velocity K amplitude of 454$\pm{109}$ km/s. When this is coupled with that of the infrared HeI absorption line (thought to originate in the WR component)  of 109$\pm{13}$  km/s (Hanson et al. 2000), it yields a mass ratio $M_{C}$/$M_{WR}$ = 0.24$\pm{0.06}$. Zdziarski et al. (2013) arrived at a similar number using the same data, but selecting a slightly different subset of the data. 
Based on the statistics of the masses and mass-loss rates of WR stars, Zdziarski et al. (2013) fixed the WR companion mass at about 10 $M_{\odot}$. This implies a mass of 2.4 $M_{\odot}$ for the compact star. We use these values in the present paper.
Zdziarski et al. (2013) estimated the mass-loss rate to be between (4.8--8.2)$\times$10$^{-6}$  $M_{\odot}$/year. We adopted the mean value 6.5$\times$10$^{-6}$  $M_{\odot}$/year, which is close to that derived from the orbital period increase
 $\dot{M}$=$M_{tot}$($\dot{P}$/2$P$)  (Ergma \& Yungelson 1998).  

Based on Gemini/GNIRS infrared spectroscopy, Koljonen and Maccarone (2017) suggested that the WR component is of spectral class WN 4-6 and that its bolometric luminosity lies between (5.5--11.5)$\times$ 10$^{38}$ erg/s. We adopted 8.0$\times$10$^{38}$ erg/s as a compromise value. For the photospheric blackbody temperature we adopted 1.25$\times$10$^{5}$ K. The WR component radius at the wind base and the binary separation are 0.95 and 3.31 solar radii, respectively.
The temperature and radius of the adopted model are close to those of the Potsdam WNE model 15-21 (in
http://www.astro.physik.uni-potsdam.de/$\sim$wrh/PoWR/WNE/). 

As the WR component is a hydrogen-deficient WN star, we adopted hydrogen, helium, and heavy element mass fractions of X=0.085, Y=0.90, and Z=0.015, respectively (Nugis \& Lamers 2000). We used solar abundances for individual heavy elements (Asplund et al. 2009). The effect on the force multiplier of using equilibrium CNO abundances is negligible. We assigned a distance of 7.4 kpc based on McCollough et al. (2016). The system parameters are collected in Table 1.  

\begin{table}
%\begin{minipage}[t]{\columnwidth}
\caption{Parameters of the Cyg X-3 binary components and stellar wind  used in this paper. Masses of the compact star  and WR star  ($M$$_{C}$ and $M$$_{WR}$) are given in $M_{\odot}$, the mass-loss rate $\dot{M}$ in  10$^{-6}$$M_{\odot}$/year, the luminosity of the WR star  $L$$_{WR}$ in 10$^{38}$ erg/s, the blackbody temperature $T_{bb}$ at the WR surface  in Kelvin degrees, the orbital period P in hours, and the distance d in kiloparsecs.}%Parameters of the Cyg 
\begin{tabular}{lcccccccc}

&&& &  & && \\
%\hline
%\\
 $M_{C}$& $M_{WR}$&$\dot{M}$&$L_{WR}$ &$T_{bb}$ &P&d\\
 2.4&10.0&6.5 &8.0 &125000&4.8&7.4\\

\end{tabular}
%\tablebib{(1) ~\citet{howarth}; (2) ~\citet{keticka}; (3)~\citet{marcolino}};
%\end{minipage}
\label{data1}
\end{table}

%\section{ Spectral states of Cyg X-3. EUV-photons between 100--230 Å}
\section{EUV photon fluxes in the spectral states of Cyg X-3}
The hybrid EQPAIR model originally developed by Coppi (1992) is suitable for modeling the broadband X-ray spectrum of Cyg X-3  (Vilhu et al. 2003; Hjalmarsdotter et al. 2009). At soft X-rays below 0.1 keV, which are not observable due to high circumstellar and interstellar absorption, the model is dominated by disk blackbody and Compton scattering from thermal electrons.

\begin{figure}
   \centering
   \includegraphics[width=9cm]{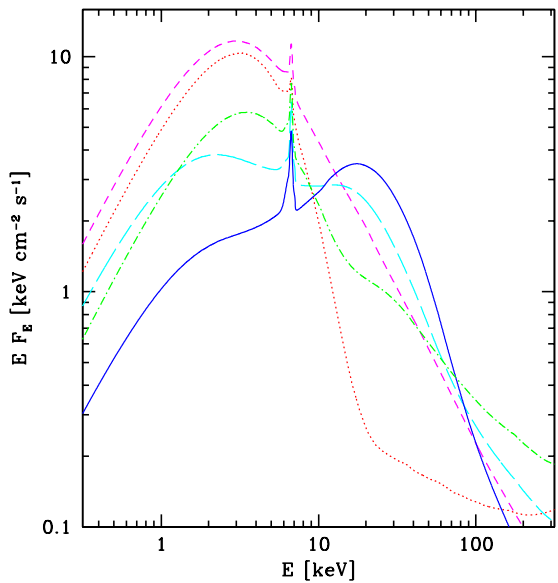}
     \caption{Unabsorbed model spectra of the five main accretion states of Cyg X-3 reproduced from Hjalmarsdotter et al. (2009). The models corresponding to  the hard state (solid blue line), the intermediate state (long-dashed cyan line), the very high state (short-dashed magenta line), the soft nonthermal state (dot-dashed green line), and the ultrasoft state (dotted red line) are shown (unabsorbed at Earth).}
         \label{spectralstates}
   \end{figure}

Using a large dataset of RXTE X-ray observations, Hjalmarsdotter et al. (2009, see their Fig. 7, reproduced here as Fig.~\ref{spectralstates}) classified the model into five main states (hard, soft nonthermal, intermediate, ultrasoft, and very high). Table A.1 gives the model fluxes at 100 eV (unabsorbed at Earth). From these fluxes and spectral shapes of Fig. 1,  the  unabsorbed number of photons between 100--230 Å (54--124 eV, where 54 eV is the HeII ionization edge) was estimated. These EUV photons are important because they decrease the size of the line force.
A similar effect is in the increase of  ionization due to X-rays (see Appendix B). The EUV flux of the hard state is uncertain due to the absorption, as noted by Zdziarski et al. (2016). 

The ionization potentials of  NiV and  FeV  (four electrons stripped off), the lines of which contribute to the line force, fall into the 100--230 Å region. As an example, photons between 70--90 eV can ionize FeV (Fe4+; see, e.g.,  El Hassan et al. 2009). If the radiator has significant flux in this region, these ions are photoionized and the line force decreases. Other smaller contributors to the line force are the lines of CIV, NV, and CrV ions, for instance, whose ionization potentials also fall in this region. For the effect of EUV photons on the line force as well as for a computation of the line force, see Appendix B.

\begin{figure}
   \centering
   \includegraphics[width=9cm]{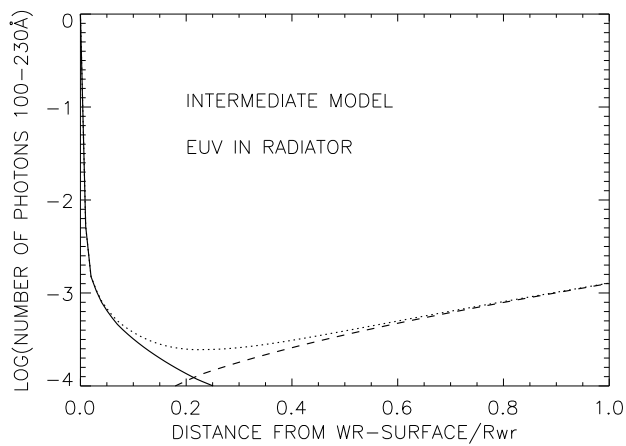}
     \caption{Number of photons  between 100--230 Å (in units of that from a 125\,000 K blackbody) as a function of  the distance from the WR star surface  (the intermediate model of Table A.1 ($f_{vol}$ = 0.3, $\Phi$ = 0) along the line joining the stars). The compact star is located at 2.4$R_{WR}$.  The solid line shows the contribution from the  WR star and the dashed line that from the  compact star (X-ray source). The dotted line shows the coadded contribution.}
         \label{xfig2}
   \end{figure}

\section{Computation of wind velocities}

\begin{figure}
   \centering
   \includegraphics[width=9cm]{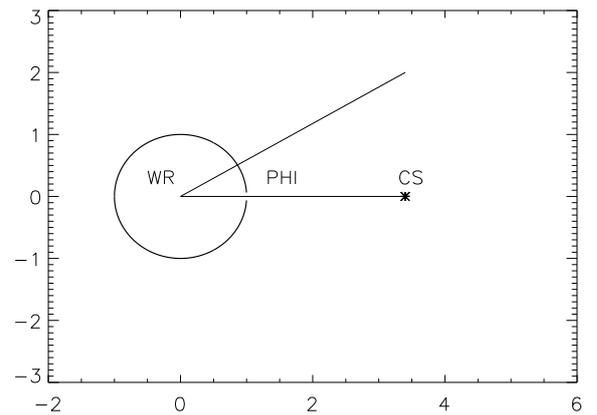}
     \caption{Geometry of the system in the orbital plane in units of the WR star  radius. The compact star (CS) is  located at distance 3.4 from the center of the WR star. The orbital angle $\Phi$ (PHI) is marked in the direction of the motion of the compact star.}
         \label{systemplot}
   \end{figure}

Wind models were computed for each of the five spectral states  taking into account the effect of X-rays from the compact star.  The wind equations were solved  along the line joining the centers  of the two stars ($\Phi$ = 0) and for different orbital angles $\Phi$ (see Fig.~\ref{systemplot}). The geometry is the same as in Krticka et al. (2012, see their Fig. 1). We also repeated their Vela X-1 computations with the present  XSTAR method, and these are outlined in Appendix C,    along with comparisons with  the CMF method for O stars from Krticka and Kubat (2017).  

Soft  X-rays from the compact star  penetrate the wind from the donor star and  change the radiation field and the number of photons in the critical EUV range 100--230 Å. In numerical  computations,  fluxes from inside (WR star) and outside (compact star) were computed  at each wind layer.  Fig.~\ref{xfig2} shows the relative  number of photons between 100--230 Å   as a function of the  distance from the WR surface (intermediate state at $\Phi$ = 0). 
The local force multiplier (line force) was then  computed using the coadded 100--230 Å fluxes  from the WR  and compact stars, respectively. Higher EUV flux corresponds to a weaker line force.
% A similar but smaller effect is the X-ray source luminosity itself. It increases the %ionization parameter $\xi$ and consequently the line force (see Appendix B). This is %dominant at distances greater than 2 stellar radii from the WR-center.

\begin{figure}
   \centering
   \includegraphics[width=9cm]{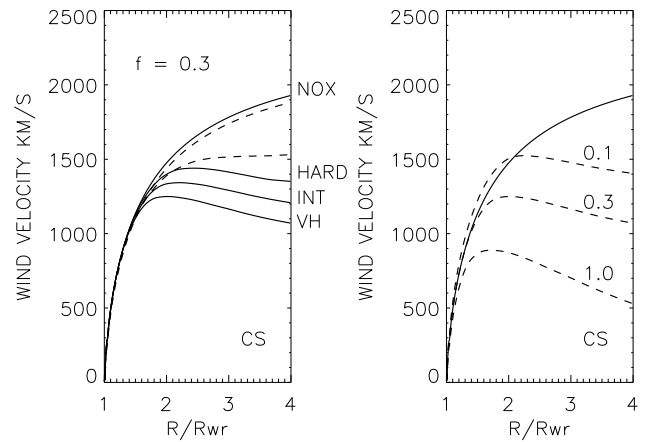}
     \caption{Left panel: Computed wind velocities  as a function of the radial distance   from the center of the WR star (in units of stellar radius)   for  the orbital angle  $\Phi$ = 0 deg (solid lines, see Fig.~\ref{systemplot}). The clumping volume filling factor $f$ = $f_{vol}$ = 0.3. VH =  very high state, INT =  intermediate state, HARD = hard state, and  NOX = the wind without X-rays from the compact star (whose distance is marked as CS in the plot). Dashed lines are for the intermediate state for angles   $\Phi$ = 30 deg (lower) and  90 deg (upper). Right panel: Wind radial velocities for the very high state with three clumping volume filling factors $f$ = $f_{vol}$ = 1.0, 0.3, and 0.1 (dashed lines). The solid line shows the wind without X-rays (NOX) and with $f$ = 0.3. The gravitational pull of the compact star is not included.}
         \label{velocities}
   \end{figure}

We considered only the supersonic part of the wind flow  (above $\delta$($r/R_{WR}$) = 0.01 from the WR surface). As shown in Fig.~\ref{xfig2}, EUV photons from the compact star do not penetrate the sonic point, only the  outer part of the supersonic wind is affected. 
At this boundary, the  wind velocity was set to 25 km/s,  slightly above the sonic point for a 10$^5$ K He gas, which is about 20 km/s. A better estimate of the sonic point properties can be found in Grassitelli et al. (2018). For a He-star  model of  10 $M_{\odot}$  and mass-loss rate of 8$\times$10$^{-6}$$M_{\odot}$/year (close  to  the Cyg X-3 value in Table 1), they  find a velocity of 19.62 km/s at the sonic point (see their Table 1). Below this boundary, the wind particle column was assumed to be 10$^{23}$cm$^{-2}$ (the whole wind column is 20 times larger).  The wind velocity scales with this column loosely as  $(column)^{0.1}$. For the model without X-rays, the assumed column gives $V_{inf}$ $\approx$ 2200 km/s (NOX in Fig.~\ref{velocities}). 

The wind was divided into 1\,000 shells with $\delta$($r/R_{WR}$) = 0.01.  The  mass-loss rate  was kept constant, and we assumed that it is determined in the subsonic part of the wind that  X-rays from the compact star  do not penetrate.  The gravitational pull of the compact star was not included.  Clumping was included assuming that all the wind mass is in the clumps and that the space between clumps is empty. In this case, the particle number density $N$  in the ionization parameter $\xi$ (Eq. B.4) is replaced by $N$/$f_{vol}$ , where $f_{vol}$ is the volume-filling factor. Clumping makes $\xi$ smaller, increasing the line force and consequently the  wind velocity (see Fig.B.4). 

The two equations which determine the flow are  mass conservation
\begin{equation}
\dot M = M_{\mathrm{{dot}}} = 4\pi  r^{2} \rho v  
\end{equation}
and  momentum balance  (neglecting  gas pressure) 
\begin{equation}
 vdv/dr = GM/r^{2} [\Gamma_{e}(1 + g_{line}  ) - 1] \, {\rm}
,\end{equation}
where the line force is 
\begin{equation}
 g_{line} =   F_{M}(r)F_{d}(r)
\end{equation}
and
\begin{equation}
\Gamma_{e} = \sigma_{e}L/(4\pi GMc)  .
\end{equation}

Using the values in Table 1 for the WR star yields $\Gamma_{e}$ = 0.34.  The value of the finite  disk  correction  factor $F_{d}(r)$  is about 0.7 at the WR surface and increases to  1 at $x$ = $r/R_{WR}$ = 1.5. It remains constant thereafter (Sundqvist et al 2014).  The force multiplier $F_{M}$ depends on the ionization parameter $\xi$, the absorption parameter $t$ (Eq. B.1), and the spectral shape of the radiator, particularly in  the EUV region around 100--230 Å.  In practice,   $F_{M}$ was interpolated from an extensive precomputed grid at each wind layer using local values of these parameters (see Appendix B).

The wind equations   were integrated numerically from $x$ = $r/R_{WR}$ = 1.01 up to 10.0 ,  resulting in velocity stratification v(r) (see Fig.~\ref{velocities}).  Without X-rays the resulting velocity stratification can be compared and fitted with a standard $\beta$-velocity model    to give the best model (see Appendix C.1). However, with  X-ray irradiation, no such standard model exists, and we used  our own formula
in which  the $\beta$-velocity model is multiplied by  a suppression factor $SF,$

\begin{equation}
 v_{model} = v_{\mathrm{{inf}}}(1 - B/x)^\beta \times SF\,
\end{equation}
\begin{equation}
%SF = 1 - slope \times (x - xkink) ,\
SF =
\begin{cases}
1 - slope \times (x - xkink), & \text{if}\ x \geq xkink \\
                        1, & \text{if}\ x < xkink
\end{cases}
\end{equation}
\begin{equation}
 B = 1.01\times[1 -  (v_{in}/v_{inf})^{1/\beta}]   
\end{equation}  
\begin{equation}
 x =  r/R_{star} ,
\end{equation}

where the $slope$ and  $xkink$ are fitting parameters.  In the inner boundary, $v_{in}$ = 25 km/s.

 The fits were improved  by integrating the wind equations  several times (without fitting), computing the EUV contamination  from the previous model, until  the average difference between two subsequent models was smaller than 1\%\ . This typically required 10--20 runs.  The results do not differ much  from the case where the X-rays were simply switched on in the NOX-model (no X-rays) and the model integrated without fitting. The computation for  $\Phi$ = 0 (along the line joining the centers of the two stars) is straightforward.  For larger angles, we numerically integrated across the wind to obtain dilution factors and absorption. Following Krticka et al. (2012), we assumed that the wind in the direction of the orbital angle $\Phi$  can be modeled by a spherically symmetric wind.

The resulting wind velocity in the NOX state  is  compatible with velocities of WN stars derived from line widths of He lines (Hamann et al. 1995).
Computations with different  orbital angles are relevant because the time in which  the wind element crosses the distance between the components (crossing time) is about 0.35--0.45 hours for our models, corresponding to orbital angles between 25--35 degrees (see Table A.1). The true wind velocity at the compact star is likely  a smear of computed velocities inside  $\Phi$ = 0--35 degrees. 

Fig. \ref{velocities} (left panel) shows the velocity stratifications for three   spectral states at $\Phi$ = 0 and moderate clumping ($f_{vol}$ = 0.3). In addition, orbital angles of 30 and 90 degrees are included for the intermediate  state. The right panel of Fig. \ref{velocities} shows velocity stratifications for the very high state with several clumping filling factors. At large distances from the WR star, the velocities approach constant values. This plot can be compared with a similar one from Krticka et al. (2018, their Fig. 4). For small clumping filling factors ($f_{vol}$ < 0.1), the wind velocity and accretion luminosity become rather insensitive to the EUV flux (see Table A1 and Fig. 5) because clumping causes the ionization $\xi$ parameter to decrease and consequently  the EUV dependence to weaken (see Fig. B.4). 

 The wind velocity around the compact star during the very high state is consistent with the P Cygni absorption of the SiXIV line at $-$900 $\pm$ 750 km/s, which is observed  most clearly at about orbital phase 0.5 (compact star in front, WR star behind; Vilhu et al. 2009). For more details on P Cygni lines and photoionization in Cyg X-3, see also Kallman et al. (2019).
 In the next section we estimate the  accretion rates and luminosities  from the computed velocities  at the compact star and compare them with those from the EQPAIR models. 
The wind velocities were computed  without the gravitational pull of the compact star to show the effect of X-rays and to apply the Bondi-Hoyle-Lyttleton model in the next section. 

\section{Accretion luminosities  during   Cyg X-3 spectral states}

 Using wind velocities at the compact star and a simple analytic wind accretion model described by Frank, King, and Raine (1992), we estimate the accretion luminosity $L_{acc}$.  This  Bondi-Hoyle-Lyttleton model  has been used widely   for instance by  Watanabe et al. (2006), Koljonen and Maccarone (2017), and Krticka et al. (2018).  A review of the topics with original references can be found in Martinez-Nunez et al. (2017). In the previous section, wind velocities were computed without the gravitational pull of  of the compact star because in the Bondi-Hoyle-Lyttleton model, the wind velocity is taken at the distance of the compact star and without its gravitational pull. This is taken into account when deriving the formula itself,  which should only be considered as a sort of scaling law.  In the model,
$L_{acc}$ strongly depends on the relative velocity $V_{rel}$,

\begin{equation}
 V_{rel}^2 = V_{wind}^2 + V_{orb}^2
,\end{equation}
 where   the orbital velocity of the compact star is V$_{orb}$ = 675 km/s using the system parameters in Table 1. The accretion luminosity is 

\begin{equation}
 L_{acc}=\dot M_{acc} \times \eta c^2 
,\end{equation}
where the accretion rate depends on the accretion radius as follows:
\begin{equation}
\dot M_{acc}/\dot M_{wind} = \pi r_{acc}^2/(4\pi a^2)
\end{equation}
\begin{equation}
 r_{acc} = 2GM_{C}/V_{rel}^2 .
\end{equation}

\begin{figure}
   \centering
  \includegraphics[width=9cm]{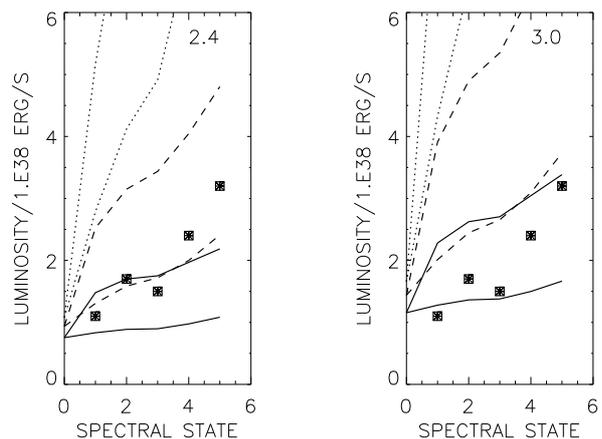}

     \caption{Left panel: Computed accretion luminosities   (in units of 10$^{38}$ erg/s) for the five spectral  states of Cyg X-3 are shown for three clumping filling factors $f$ = $f_{vol}$ = 1.0 (dotted lines), = 0.3 (dashed lines), and = 0.1 (solid lines).   The orbital angle  $\Phi$ = 0 deg  for the upper curves and  $\Phi$ = 30 deg for the lower ones. The filled squares show the EQPAIR model luminosities  ($L_{bol}$ of Table A.1) The compact star mass is 2.4  $M_{\odot}$. In the abscissa 0 indicates  NOX (no X-rays), 1 hard, 2 soft nonthermal, 3 intermediate, 4 ultrasoft, and  5 very high state. Right panel: Same as in the left panel, but the compact star mass is 3.0  $M_{\odot}$. 
}
         \label{lx38rel}
   \end{figure}

 The mass-to-energy conversion factor $\eta$ = 0.1 was used, which is approximately  valid for neutron stars and black holes
(see, e.g.,  Frank, King and Raine 1992).
Accretion luminosities were  calculated  for the five spectral  states using the computed wind velocities at the compact star (see Fig.~\ref{velocities} and Table A.1) for three clumping filling factors  $f_{vol}$ = 1.0 (no clumping), 0.3, and 0.1.   These are shown  in Fig.~\ref{lx38rel}. The filled squares in Fig.~\ref{lx38rel} show the EQPAIR model luminosities  (L$_{bol}$ of Table A.1). 

The left panel of  Fig.~\ref{lx38rel} shows the luminosities computed for the compact star mass  2.4  $M_{\odot}$  , and  the right panel shows the luminosities for a higher compact star   mass   3.0  $M_{\odot}$. Because the realistic wind velocity probably lies between  orbital angles 0 and 30 degrees (30 deg is close to the crossing times in Table A.1),  moderate clumping  favors  2.4  $M_{\odot}$ and  higher clumping  favors  3.0  $M_{\odot}$.  Lower masses require less clumping.   At very high clumping,  the correlation between different spectral states disappears. 

Fitting the mean  bolometric luminosity value of spectral states ($L_{bol}$=2$\times$10$^{38}$ erg/s) with similar means from our computations at orbital angle $\Phi$ = 15 degrees, we derived the relation between the clumping filling factor and compact star mass ($f_{vol}$, $M_{C}$): (0.70,1.4), (0.28,2.0), (0.16,2.4), and (0.08,3.0). Szostek and Zdziarski (2008) estimated  values between 0.07 - 0.3 for the one-phase filling factor (used in the present work).  This points to a compact star mass between (2-3)$M_{\odot}$.

%\end{document}
\section{Discussion}

We computed the force multiplier (line force)   using the CAK method, and the wind was approximated by the photoionized plasma code XSTAR (Stevens and Kallman 1990; Kallman  2018).
XSTAR is designed to calculate the ionization and thermal balance in gases exposed to ionizing radiation. It contains a relatively complete collection of relevant atomic processes and corresponding atomic data. It does not by itself include processes associated with wind flows, such as shock heating, adiabatic cooling, or time-dependent effects. It also has a simple treatment of radiation transport that provides approximate results  in situations involving truly isotropic radiation fields or line blanketing, for example. 

Key simplifications employed by XSTAR are with regard to the radiative transfer solution.  Escape of line radiation is treated using an escape probability formalism, and this affects the temperature through the net radiative cooling.  Transfer of the ionizing continuum is treated using a single-stream integration of the equation of transfer, including opacities and emissivities calculated from the local level populations.  Thus, there is no allowance for the inward propagation of diffusely emitted radiation. 

 This last approximation is likely to be most important for the results presented here because strong winds can have large optical depths in the ionizing continua of H and He II.  It would be desirable to verify our current results for the force multiplier using a transfer solution that does not have this limitation.   Limitations of the escape probability assumption for line escape have been pointed out by Hubeny (2001), for example.  However, for hot star winds, the strong velocity gradient reduces most line optical depths to approximately a few at most, and so the lines are effectively thin and the radiative cooling is not strongly affected. However, the simplifications made do not affect the EUV sensitivity of the force multiplier, which is the key issue here.

The assumed  masses for the components may represent  just one possible solution for Cyg X-3. A more massive possibility within errors is 3.0 + 10  $M_{\odot}$, for example, yielding 50\%\ higher computed accretion luminosities. A less massive alternative    2.0 + 10  $M_{\odot}$ yields  the opposite. A tolerable match with the EQPAIR models can be provided by moderate clumping  ($f_{vol}$ = 0.1 - 0.3)  and $M_{C}$ =  2.4  $M_{\odot}$ or by $f_{vol}$ = 0.1 and $M_{C}$ =  3.0 (see   Fig.~\ref{lx38rel}). Smaller compact star masses require less clumping, which is an unknown parameter in this study. However, typical volume filling factors for WR-stars have been estimated to be 0.1 - 0.25 (e.g., Hamann and Koesterke 1998).

The region  in which the photoionization of the main contributors to the line force takes place was assumed to be 100 Å--230 Å. However, this region could be shallower, for instance, 100--177 Å, because the ionization potentials of the main contributors NiV and FeV are at 164 Å.  In this case,  the computed accretion luminosities   would  increase (by a factor of 1.5).    One possible caveat is that the estimated EUV fluxes from the EQPAIR models do not necessarily reflect reality. In a sense, the EUV flux is always model dependent because we cannot obtain any observations in these wavelengths.

Accretion disks are frequently assumed in the context of wind accretion. Accretion disk formation in radiatively driven wind accretion has recently been studied by Karino et al (2019). A good illustration of the Cyg X-3 disk can be found in Hjalmarsdotter et al. (2009).
The origin of transitions  between the different states probably lies in some form  of disk instability. In the hard state, the radius of the inner accretion flow is large (and  the EUV flux   low). In the soft states, the inner disk boundary  shrinks, and consequently, the EUV flux  increases, the wind velocity decreases, and  the accretion luminosity $L_{acc}$ increases (see Fig. 7 in Hjalmarsdotter et al. 2009 and discussion therein). 

The disk and  wind may interact. This means that the changing EUV flux   from the compact star might account for the different luminosity and spectral states of Cyg X-3. When the EUV flux is high, the X-ray accretion luminosity is also high, and vice versa.  Our computations together with the EQPAIR model luminosities favor this picture  if both the compact star mass and clumping are suitable (see  Fig.~\ref{lx38rel}). 
We might speculate that during an eventual extremely high state (flare)  the computed wind  at the compact star may stop, the X-ray production ends, and the wind returns to the NOX state. During this state, the accretion luminosity does not differ much from the hard-state luminosity (see  Fig.~\ref{lx38rel}).
The hard state may thus be the most natural and long-lasting state of Cyg X-3.

Antokhin and Cherepashchuk (2019)  extended the list of times of X-ray minima and fit  the $O-C$ diagram by a sum of quadratic and sinusoidal (with period 15.79 years) functions. They suggested the existence of either apsidal motion or a third body.  The mass loss, causing the period increase, is behind the overall shape of the $O-C$ diagram. For the  quadratic plus sinusoidal ephemerides, Antokhin and Cherepashchuk find  $\dot{P}$ = 5.629$\times$$10^{-10}$ , corresponding to a relative period change  of $\dot{P}$/$P$ = 1.029$\times$$10^{-6}$/year. $\dot{P}$ is related to the coefficient $c$ of the quadratic term by $c$ = $P$$\dot{P}$/2.  We estimate the change of $\dot{P}$/$P$ in our models  below. 

For a nonconservative mass transfer  by fast isotropic wind (Jeans mode, see Ziolkowski and Zdziarski 2018, their eq. 15 with $\alpha$=1): 
\begin{equation}
\frac{ \dot P}{P}=  \frac{\dot M_{wind}}{M_{WR}}[3 - 3\beta \frac{M_{WR}}{M_{C}}-(1-\beta)\frac{M_{WR}}{M_{tot}}-3(1-\beta)\frac{M_{C}}{M_{tot}}]
%\beta = \dot{M}_{acc}/\dot{M}_{wind}
,\end{equation}

where $\beta$ is the fraction of the wind accreted by the compact star  $\dot{M}_{acc}$/$\dot{M}_{wind}$.

 For an acceptable model (10$M_{\odot}$ + 2.4$M_{\odot}$, $f_{vol}$ = 0.16, see Sect. 5), we have $\beta$ = 0.002 and 0.005 for the hard state and the very high state, respectively. For the period change, these give  $\dot{P}$/$P$ values 1.034$\times$$10^{-6}$/year (hard) and 1.012$\times$$10^{-6}$/year (very high), which differ by 2\%. Hence, the effect of spectral states on the $O-C$ diagram is  very small through the quadratic term coefficient. Moreover, in the $ASM/RXTE$ data, Cyg X-3 spent  50\%\ of the time in the low state (< 10 cps) and 22\%\ in the high state (> 20 cps). Hence, the average of $\dot{P}$/$P$  is closer to the hard-state value as observed.  Locally, like during and after  the jigsaw pattern detected  in the $ASM/RXTE$ data  between MJD = 50700 - 51600  (when Cyg X-3 was mostly in the low hard state), this effect is unobservable.  The jigsaw pattern is probably  caused by the temporal irregular distribution of $ASM$ data, as suggested by Antokhin and Cherepashchuk.

\section{Conclusions}
We constructed wind models in the supersonic part   of the WR component of Cyg X-3  using  force multipliers (line force) computed  with CAK formalism and XSTAR physics (following the pioneering work of Stevens and Kallman 1990; for the XSTAR-code, see Kallman 2018). The line force strongly depends on the EUV flux of the radiator  (100--230 Å), to which the X-rays from the compact star contribute. 
The main contributors are lines of  NiV and FeV   ions (to a lesser extent, e.g., CIV, NV, and CrV). Their (photo)ionization potentials fall  between 100--230 Å (54--124 eV).  In this region  the radiation field controls these lines and consequently the size of the force multiplier. Higher EUV flux from the compact star means a weaker line force, which in turn leads to a lower wind velocity.    

 Because  X-rays do not reach the  subsonic part of the wind, we assumed that the mass-loss rate is unaffected by X-rays. We included  clumping  by assuming that all the wind mass is in the clumps. We modeled the wind  at different orbital angles    (see  Fig.~\ref{systemplot}). Computations with different orbital angles are relevant because the time in which  the wind element crosses the distance between the components (crossing time) is about 0.35--0.45 hours in our models. This corresponds to orbital angles between 25--35 degrees (see Table A.1).  

Radiation from the compact component  increases the EUV flux and diminishes the wind velocity (Fig.~\ref{velocities}). This enhances the  accretion rate, and consequently, the accretion luminosity (Fig.~\ref{lx38rel}).  The computed Bondi-Hoyle-Lyttleton accretion luminosities $L_{acc}$ correlate moderately well with the  luminosities of the five main spectral states of Cyg X-3 defined by Hjalmarsdotter et al. (2009), provided that the clumping is moderate ($f_{vol}$ = 0.1 - 0.3) and the mass of the compact star is about (2 - 3)$M_{\odot}$ (with the WR-star mass 10$M_{\odot}$, see Fig.~\ref{lx38rel}).  For higher clumping ($f_{vol}$ << 0.1) the EUV sensitivity disappears because clumping decreases the ionization parameter $\xi$  and the line force dependence on EUV flux (see  Fig.~\ref{xfig1}).

We compare our  XSTAR method  with the CMF method (comoving frame) in Appendix C. The results are qualitatively similar for both O stars (Krticka and Kubat, 2017) and X-ray suppression in Vela X-1 (Krticka et al. 2012).

\begin{acknowledgements}
  We thank Drs. Hjalmarsdotter and Krticka for permission to use their Figures (Fig.1 and Fig.C.3) and Drs Andrzei Zdziarski and Linnea Hjalmarsdotter for valuable comments. K.I.I.K was supported by the Academy of Finland projects 320045 and 320085.
\end{acknowledgements}

\begin{appendix}

\section{Computed wind velocities}

 In Table we list A.1 the EUV/EQPAIR fluxes and bolometric luminosities of Cyg X-3 spectral states (Hjalmarsdotter et al. 2009) and the computed  wind velocities at the compact star (this study).

\begin{table*}
%\begin{table*}[t]
\centering
%\begin{minipage}[t]{\columnwidth}
\caption{Luminosities and EUV-fluxes of the five basic states of Cyg X-3 from RXTE observations and EQPAIR models (Hjalmarsdotter et al. 2009, first two rows) and  computed wind velocities at the compact star   (this  study, last  rows). NOX is the state without X-rays. $L_{bol}$ (in units of 10$^{38}$ erg/s) is the EQPAIR model luminosity. EFE (E$\times$F$_{E}$; in units of KeV/cm$^{2}$/s) is the unabsorbed flux at Earth (at 100 eV, 124 Å).   Computed wind velocities at the compact star Vwind (km/s)  are given for two orbital  angles $\Phi$ = 0 and 30 degrees  (the orbit = 360 deg, see Fig.~\ref{systemplot}) and for three clumping filling factors $f_{vol}$ = 1.0 (no clumping),  0.3, and  0.1. Time (in degrees,  the orbital period  = 360 deg) is the crossing time of a wind particle from the WR surface to the compact star distance.}
% Vwind is the wind velocity at the compact star (X-source) from which the accretion l%uminosity Lacc (last row) was computed. Vinf is the iterated model velocity at %infinity and Chi2 is the fitting $\chi^2$ (DOF=239).  Values used for the WR-%companion are given in the last column. }
%end{document}
%\renewcommand{\footnoterule}{}  % to avoid a line before footnotes
\begin{tabular}{lcccccccc}

&PARAMETER&NOX &HARD  &SOFT NON-TH & INTERMEDIATE  & ULTRASOFT&VERY HIGH \\
%\hline
%\\
& $L_{bol}$ & &1.1&1.7&1.5&2.4&3.2\\
& EFE & &0.03&0.07&0.11&0.14&0.19& \\
& $f_{vol}$ = 1.0 : & & & & & & \\
&Vwind  $\Phi$=0&1779 &1094 &933  & 841 &747  &627 \\
&Vwind  $\Phi$=30&1779 &1340  &1181  &1113  &983 &885   \\
&Time  $\Phi$=0&29&35&38&40&43&47 \\
%&Lacc  $\Phi$=0&0.82 &3.6 &4.7 &6.0 & 6.7&8.2 \\
%&Lacc  $\Phi$=30& &2.2 &2.8 &3.6 &4.0 &4.9 \\
& $f_{vol}$ = 0.3 : & & & & & & \\
&Vwind  $\Phi$=0&1851  &1382  &1289  &1253  &1188 &1121  \\
&Vwind  $\Phi$=30&1851  &1682  &1590  &1552  &1483  &1401 \\
&Time  $\Phi$=0&28&30&31&31&32&33 \\
& $f_{vol}$ = 0.1 : & & & & & & \\
&Vwind  $\Phi$=0& 1967 &1622  &1557  &1543 &1490  &1444  \\
&Vwind  $\Phi$=30&1967  &1913  &1879  &1873  &1830  &1776 \\
&Time  $\Phi$=0&25&26&26&27&27&28 \\
%&Vwind  $\Phi$=30& &2010 &1870 &1720 &1680 &1580  \\
%&Lacc  $\Phi$=0&0.50 &1.2 &1.6 &2.0 &2.2 &2.7 \\
%&Lacc  $\Phi$=30& &0.7 &0.9 &1.2 &1.3&1.6  \\
\end{tabular}
%\tablebib{(1) ~\citet{howarth}; (2) ~\citet{keticka}; (3)~\citet{marcolino}};
%\end{minipage}
\label{data}
\end{table*}
%\end{appendix}

%\begin{appendix}

\section{ Methods for computing the line force (force multiplier). Effect of EUV photons.}
The force multiplier (line force)  $F_{M}$ was computed following the method described by CAK. An important difference is that CAK (and other compilations, e.g., Gayley et al. 1995) assume that the ionization and excitation in the wind is given by  the Saha-Boltzmann distribution modified by an analytic dilution factor applied to all elements.  We computed the ionization distribution in the wind by balancing the ionization and recombination due to the stellar radiation field.  We also employed the Boltzmann distribution to determine the populations of excited levels.    The radiation field was computed using a single-stream integration outward from the stellar photosphere.  In this way, we calculated the line force appropriate to any spectral form of the photospheric spectrum and at any layer in the wind. This permitted us to integrate wind equations from the surface to infinity  in a self-consistent manner because the upward force is known.

The ionization balance and outward transfer of photospheric radiation were calculated using the  XSTAR photoionization code  (Stevens \& Kallman, 1990; Kallman, 2018).  The ion fractions at each spatial position in the wind were used to calculate the force multiplier $F_{M}$ by summing the CAK line force expression over an ensemble of lines.  The list of lines was  taken from Kurucz (http://kurucz.harvard.edu/linelists.html) and permitted us to use local wind parameters. The line list is more extensive than in the original CAK work.  It is assumed that the wind is spherically symmetric around the donor star (the ionizing source). 

XSTAR  calculates the ionization balance for all the elements with atomic number Z $\leq$ 30 together with  the radiative equilibrium temperature. 
It calculates full nonlocal thermal equilibrium (NLTE) level populations for all the ions of these elements.  It includes a fairly complete treatment of the level structure of each ion, that is, more than $\sim$50 levels per ion, and up to several hundred levels for some ions.  Many relevant processes affecting level populations are included such as radiative decays, electron impact excitation and ionization, photoionization, and Auger decays.   The electron kinetic temperature is calculated by imposing a balance between heating from fast photoelectrons and Compton scattering with radiative cooling.

 For a fixed radiator, the force multiplier $F_{M}(t, \xi, N)$ is  a three-dimensional function of  the absorption parameter $t$, ionization parameter $\xi$ (erg/s$\times$cm), and particle number density $N$ ($cm^{-3}$),
\begin{equation}
  t = \sigma_{\mathrm{e}}v_{\mathrm{th}}\rho(dv/dr)^{-1} 
\end{equation}
\begin{equation}
    \xi = \frac{L_{\mathrm{ly}}}{Nr^{2}}\ .
\end{equation} 

Here  $L_{ly}$ is the  radiation  luminosity below  the hydrogen ionization limit 912 Å,  $\sigma_{e}$ is the electron scattering coefficient, $\rho$ and $N$ are the gas   and particle number density, respectively, $v$ is the outward wind velocity, $r$ is the radial distance from the ionizing source, and $v_{th}$ is the gas thermal velocity (typically $\sim$10 km/s).  The dependence on   $N$ is weak (Vilhu and Kallman 2019). The force multiplier was computed  assuming a point source radiator.  When applied to stars,  the dilution factor $r^{-2}$ in  $\xi$ (Eq. B2) should be replaced by  the finite disk factor (Mewe and Schrijver 1978),

\begin{equation}
 W(r) = 0.5[1 - \sqrt(1 - (R_{star}/r)^2)]R_{star}^{-2} .
\end{equation}

In our paper the radiator consists of two sources: the WR-star and the (point source) compact star.  The ionization parameter was computed taking the (absorbed)  contributions $L_{lywr}$ and $L_{lyx}$ from both sources into account,

\begin{equation}
   \xi = \frac{L_{\mathrm{lywr}}}{N}W(rwr)   + \frac{L_{\mathrm{lyx}}}{N(rx)^{2}}\ , 
\end{equation}

where $rwr$ and $rx$ are the distances from the WR center and compact star, respectively. $L_{lywr}$ and $L_{lyx}$ are component luminosities below the Lyman edge at a specific point in the wind, absorbed along the path from the WR-surface and the compact star (X-ray source), respectively (by exp(-absorption$\times$column)). X-rays increase the ionization parameter $\xi$ through the Lyman luminosities and diminish the line force. In the case of Cyg X-3, the EUV-photons from the compact star are more important. Their number effectively controls the line force.

The chemical abundances of the wind and the ionizing source spectrum  can be specified as input data.  The method is in principle  the same as in the classical work of CAK. However,  we explicitly included $\xi$ and $t$  and computed it locally in the wind.  In the case of clumping, with volume-filling factor $f_{vol}$ and all mass in the clumps, the particle number density $N$ in Eq. B2 should be replaced by $N$/$f_{vol}$.

The force multiplier depends on $t$ and on the cross section for line absorption,
\begin{equation}
$$F_{M}(t,\xi)=\Sigma_{lines}{\frac{\Delta\nu_DF_\nu}{F} \frac{1}{t}(1-e^{\eta t})}$$
,\end{equation}

where $\Delta\nu_D$ is the thermal Doppler width of the line, $F$ is the total flux in the continuum, $F_\nu$ is the monochromatic flux at the line energy, and 
\begin{equation}
$$\eta=\kappa_{line}/\sigma_e$$
\end{equation}
\begin{equation}
$$\kappa_{line}=\frac{\pi e^2}{m_ec} g_L f \frac{N_L/g_L-N_U/g_U}{\rho\Delta\nu_D}$$
,\end{equation}
where $N_L$ and $N_U$ are the lower and upper level populations, $g_i$  are the statistical weights, and $f$ is the oscillator strength.  Our code (xstarfmult.f) uses the ion fractions calculated as described above to sum over lines to calculate $F_{M}(t, \xi)$.  Ground-state level populations were taken directly from XSTAR;  excited level populations  were calculated  assuming a Boltzmann (LTE) distribution.  This is necessitated by the  use of the  Kurucz line list, which does not include collisional rates or other level-specific quantities that would allow a full NLTE calculation of populations.  Most previous calculations of the CAK force multipliers have employed the LTE assumption for excited levels as well (e.g., CAK; Abbott 1982;  Gayley 1995).

 $F_{M}$ grids were computed for the chemical composition of WNE stars for many $\xi$- and $t$-values and radiators around  125 000 K blackbody  with different amounts of flux cut below 230 Å. Cut blackbodies give the same size force multipliers as realistic absorbed models (Vilhu and Kallman 2019, their Figs. 2 and 3).  In the course of  computations, at a fixed layer in the wind, the real absorbed radiator spectrum was compared with the grid spectra  and   $F_{M}$ interpolated together with local $\xi$  and $t$ values. The EUV contamination of the X-ray source was  included. 

\begin{figure}
   \centering
   \includegraphics[width=9cm]{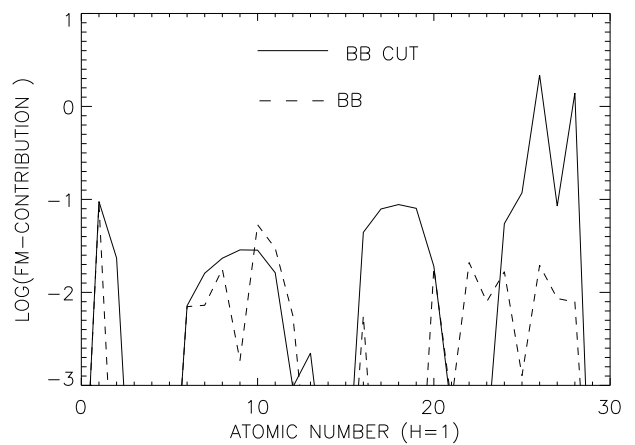}
     \caption{Contributions to the force multiplier ($F_{M}$) as a function of the element atomic number (26 for Fe). A 135 000 K blackbody (dashed line) and  below 230 Å totally cut blackbody (Nphot = 0 between 100--230 Å) with the same temperature (solid line)  were used as  radiators, and the force multiplier was calculated at log($\xi$) = 2.5 and log($t$) = $-$1.84.  The y-axis has a logarithmic scale.  }
         \label{elemcontri}
   \end{figure}

\begin{figure}
   \centering
   \includegraphics[width=9cm]{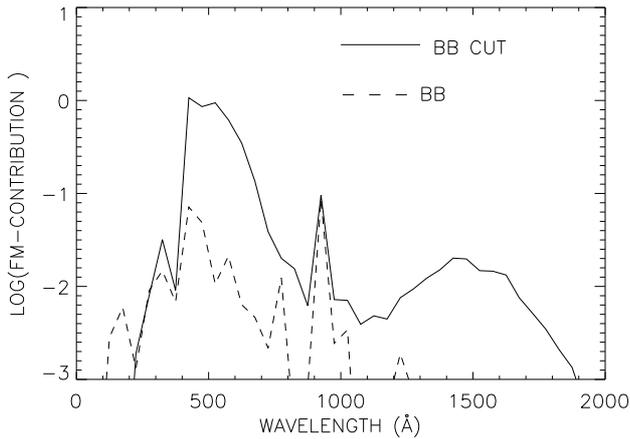}
     \caption{Same as in Fig.~\ref{elemcontri}, but contributions to the force multiplier counted inside  50 Å wide bands are shown as a function of wavelength.    }
         \label{wlcontri}
   \end{figure}

\begin{figure}
   \centering
   \includegraphics[width=9cm]{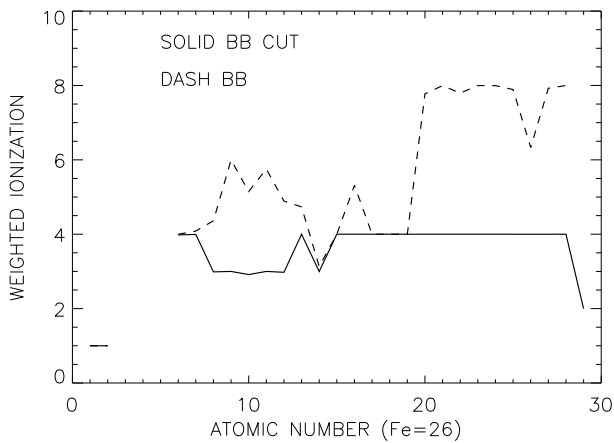}
     \caption{Mean ionization (weighted by the contribution to the force multiplier) as a function of  atomic number shown for   log($\xi$) = 2.5 and log($t$) = $-$1.84 (as in Figs. \ref{elemcontri} and  \ref{wlcontri}). }
         \label{ioncontri}
   \end{figure}

The lines of NiV and  FeV  ions  are the strongest contributors to the force multiplier (in addition to, e.g., CIV, NV, and CrV). The lines are mainly in the Lyman-continuum below 912 Å. When these ions ionize, the lines disappear, and consequently, the force multiplier becomes smaller.  This ionizing takes place if the radiator has significant flux below 230 Å .  
Figures  \ref{elemcontri} and  \ref{wlcontri} show this effect for a 135\,000 K blackbody (slightly differing from 125 000 K of Table 1) as a function of element atomic number and wavelength.  Fig.  \ref{ioncontri} shows the mean ionization weighted by the contribution to the force multiplier (4 means that four electrons are stripped off). The soft X-ray suppression clearly vanishes when flux below 230 Å is absorbed, which increases the contribution of the iron group elements. This property of force multiplier also helps to explain the momentum problem  of  WR winds (Vilhu and Kallman 2019).

\begin{figure}
   \centering
   \includegraphics[width=9cm]{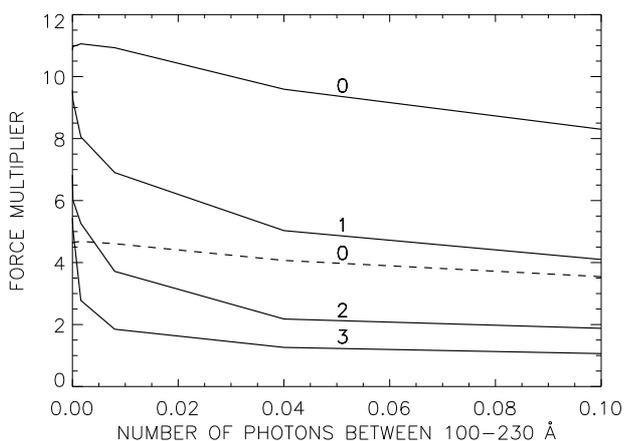}
     \caption{Force multiplier vs. the number of photons in the radiator between 100--230 Å  (in units of that in 125\,000 K blackbody).  Curves for four log($\xi$)-values (0, 1, 2, 3) with log($t$) = -1.85  are plotted (solid lines). The dashed line is for the (log($\xi$), log($t$))-pair (0,-1.36). }
         \label{xfig1}
   \end{figure}

\begin{figure}
   \centering
   \includegraphics[width=9cm]{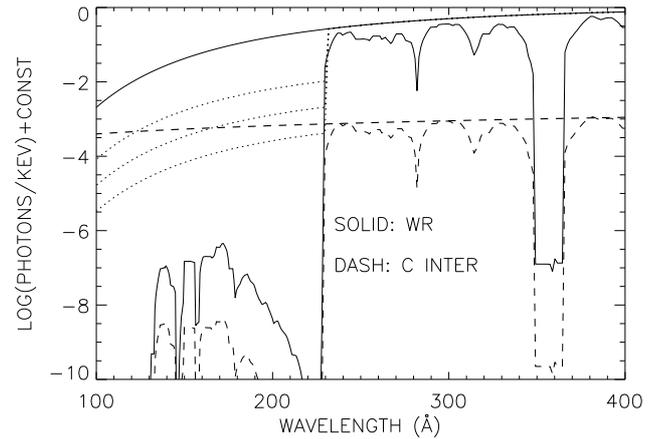}
     \caption{Spectral energy distributions for the radiators of the intermediate spectral state at $\Phi$ = 0 (along the line joining the centers of components). Solid lines show the WR spectra at the photosphere (125\,000 K blackbody) and the absorbed spectra at distance 2$R_{WR}$ from the WR surface. The dashed lines show the compact star spectra (unabsorbed and absorbed at distance 0.1$R_{WR}$ from the WR surface). The dotted lines show some partially cut blackbodies used to compute force multiplier grids. }
         \label{xfiguusi}
   \end{figure}

 Fig.~\ref{xfig1} shows the change in the force multiplier as a function of the number of photons in the 
radiator between 100--230 Å (relative to that of a 125\,000 K blackbody).  The plot shows curves for several (log($\xi$), log($t$)) pairs. Decreasing  $\xi$ or $t$ enhances the line force.  Fig.~\ref{xfiguusi} shows some spectral energy distributions for the radiators of the intermediate spectral state at $\Phi$ = 0 (along the line joining the centers of components) and some examples of partially cut blackbodies with which force multiplier grids were computed. At each wind layer, the coadded number of photons (from WR and compact star)  between 100--230 Å  was computed and the force multiplier  was interpolated  from a precomputed grid. 

\section{Comparison of our method (XSTAR) and  CMF methods} 
\subsection{OB-star winds}
 
Krticka and Kubat (2017) computed comoving frame winds (CMF) for a  model grid of O stars comprising main-sequence stars, giants, and supergiants (their Table 1).  Using  the masses, radii, effective temperatures, and solar chemical abundances (Asplund et al. 2009), we computed the  wind models with the XSTAR method. The data were reproduced from extensive computations in Vilhu and Kallman (2019).
The wind equations   were integrated numerically,  resulting in velocity stratification v(r)  that was compared with the widely used $\beta$-velocity model, 
\begin{equation}
 v_{model} = v_{\mathrm{{inf}}}(1 - B/x)^\beta \,
\end{equation}
%\end{document}
\begin{equation}
 B = 1. -  (v_{in}/v_{inf})^{1/\beta}   
\end{equation}  
\begin{equation}
 x =  r/R_{star} .
\end{equation}

The integrations were started above the subsonic region at $x$ = 1.01   and ended at  $x$  = 10,  using 1000 grid points with  $\delta x$ = 0.01. The subsonic part of the wind enters only through the boundary condition  $v_{in}$ = 10 km/s (at x = 1.0). 
The resulting $v(x)$ was then compared with the model $v_{model}(x)$ . The unknown free parameters mass-loss rate $M$$_{dot}$ and $v_{inf}$ (velocity at infinity)  were  iterated using the IDL procedure mpcurvefit.pro (written by Craig Markwardt) to obtain ($v(x) - v_{model}(x)$)/$v_{model}(x)$  to approach  zero at each $x$.  The procedure mpcurvefit.pro  performs a Levenberg-Marquardt least-squares fit, and no weighting was used.   Density clumping was not used by either Krticka and Kubat (2017) or here. 
The formal one-sigma errors of each parameter were computed from the covariance matrix.

\begin{figure}
   \centering
   \includegraphics[width=9cm]{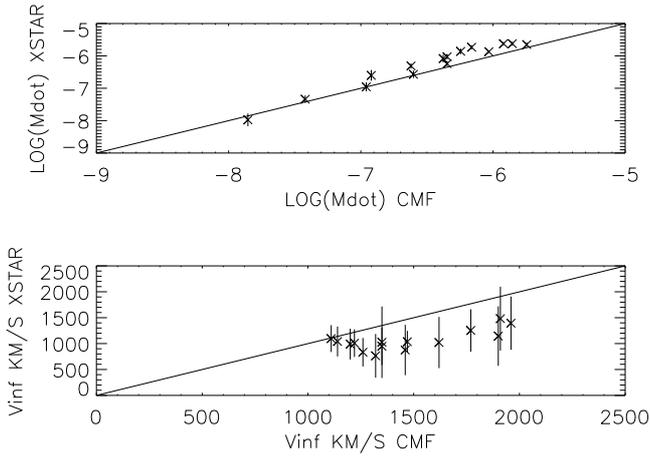}
     \caption{Comparison of CMF models of Krticka and Kubat (2017) with XSTAR models of our study.   Mdot (solar masses/year) and  $v_{inf}$ (km/s) are the mass-loss rate and velocity at infinity, respectively.}         \label{cmfvsxstar}
 \end{figure}

$\beta$ was set to 0.6, giving overall the best-fitting results. Using  $\beta$= 1.0,   $\dot{M}$ and  $v_{inf}$  should be multiplied on average by 0.7 and 1.4, respectively (see details in Vilhu and Kallman, 2019). Fig. \ref{cmfvsxstar} compares our XSTAR models and the CMF models.  Increasing  $\beta$ from 0.6 to 0.8 would give better agreement at high mass-loss rates. However, the overall agreement is moderately good.
\subsection{X-ray contaminated wind of Vela X-1.}
 Krticka et al. (2012) applied the CMF method to the X-ray contaminated wind of Vela X-1 pulsar. This is a binary system with an orbital period of 8.96 days and a separation of 53.4 $R_{\odot}$. The masses of the neutron star and of the companion O star are 1.88 and 23.5 solar masses, respectively.  The O star has an effective temperature of 27 000 K and a radius 30  $R_{\odot}$ , and the wind mass-loss rate is 1.5$\times$ 10$^{-6}$ $M_{\odot}$/year. The X-ray luminosity of the neutron star 3.5$\times$ 10$^{36}$ erg/s was assumed  with  a photon index $\Gamma$ = 1 (Watanabe et al. 2006). 
Force multiplier grids with different EUV contaminations were computed for the 27\,000 K blackbody in a similar fashion as explained for Cyg X-3 (Appendix B).  In the Vela X-1 case, the number of EUV photons relative to that of 27\,000 K blackbody is larger than 1.
 The geometry is illustrated in Fig. 1 of Krticka et al. (2012).  

\begin{figure}
   \centering
   \includegraphics[width=9cm]{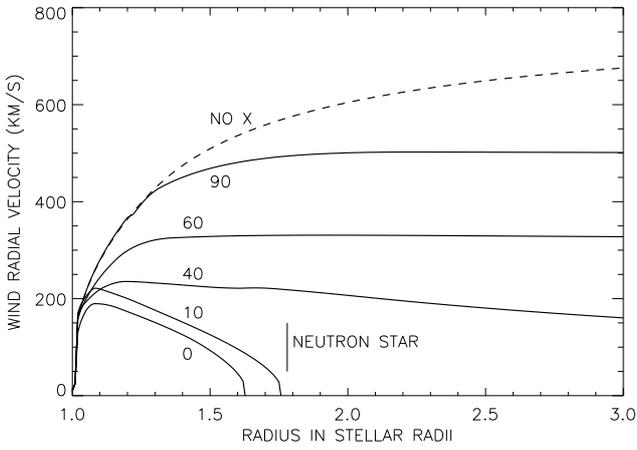}
     \caption{Radial wind velocity stratifications of Vela X-1 for different orbital angles $\Phi$  marked (0--90 degrees).  The label 'NO X'  shows the wind without X-rays from the neutron star at radius 1.78. The radius is measured from the center of the O star in units of  30  $R_{\odot}$ (= O-star radius) }
         \label{vela}
 \end{figure}

\begin{figure}
   \centering
   \includegraphics[width=9cm]{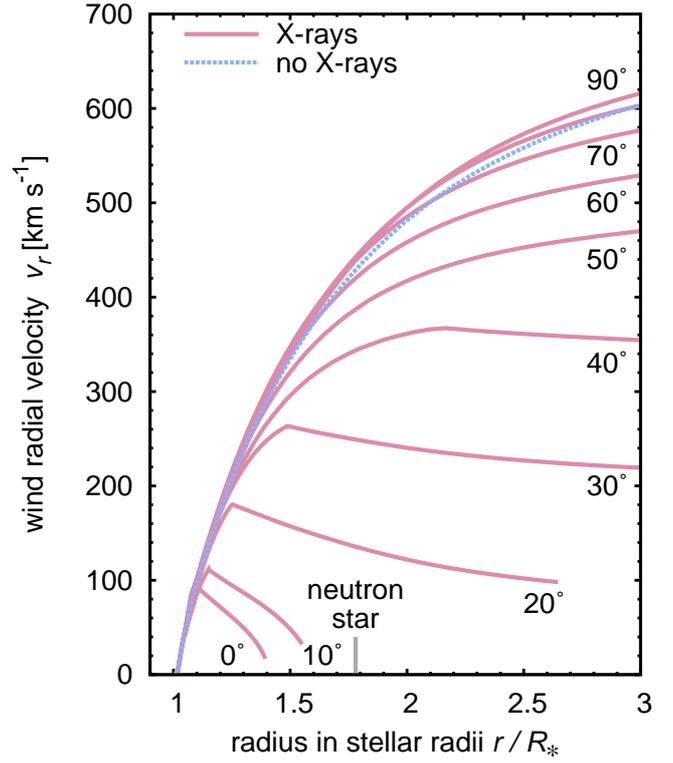}
     \caption{Related Vela X-1 study  reproduced from Krticka et al. (2012). Radial wind velocity stratifications for different orbital angles $\Phi$. The position of the neutron star is marked in the graph.}
         \label{phix}
 \end{figure}

Our method was applied to Vela X-1 in a similar fashion as to Cyg X-3. Fig.~\ref{vela} shows the computed velocity stratifications for different  orbital angles $\Phi$.     The trend is similar to that in Krticka et al. (2012, their Fig. 3, reproduced here as Fig.~\ref{phix}). Our study gives a weaker   $\Phi$-dependence, but the results are qualitatively similar. Quantitative differences are expected because the approaches are very different.

%\begin{figure}
  % \centering
  % \includegraphics[width=9cm]{vela2.eps}
   %  \caption{vela2 }
      %   \label{vela2}
 %\end{figure}
\end{appendix}
\end{document}